\documentclass[aps,pre,floatfix,twocolumn]{revtex4} 

\usepackage{natbib}
\usepackage{graphicx}

\begin{document} 

\title{Adaptive Coevolutionary Networks -- A Review} 

\author{Thilo Gross}
\affiliation{Max-Planck Intitut f\"ur Physik komplexer Systeme, Biological Physics Section, N\"othnitzer Stra\ss e 38,
01187 Dresden, Germany. Email: thilo.gross@physics.org}
\author{Bernd Blasius} 
\affiliation{Institute for Chemistry and Biology of the Marine Environment (ICBM),
Carl von Ossietzky Universit\"at, 26111 Oldenburg, Germany. Email: blasius@icbm.de}

\date{\today} 

\begin{abstract}
Adaptive networks appear in many biological applications. They combine topological 
evolution of the network with dynamics in the network nodes.  
Recently, the dynamics of adaptive networks has been investigated in a number
of parallel studies from different fields, ranging from genomics to game theory. 
Here we review these recent developments and show that they can be viewed from a unique angle. 
We demonstrate that all these studies are characterized by common themes, 
most prominently: complex dynamics and robust topological self-organization 
based on simple local rules. 
\end{abstract}

\maketitle


\section{Introduction}
Complex networks are ubiquitous in nature. They occur in a large variety of real-world
systems ranging from ecology and epidemiology to neuroscience, socio-economics and
computer science \cite{Barabasi:Review,Dorogovtsev:Review,Newman:Review,Newman:NetworkBible}. While physics 
has for a long time been concerned with well-mixed systems, lattices and spatially explicit
models, the investigation of complex networks has in the recent years received a
rapidly increasing amount of attention. In particular, the need to protect or optimize 
natural networks as well as the goal to create robust and efficient technical nets, prove to be 
strong incentives for research.

\begin{figure}
\includegraphics[width=2.5in]{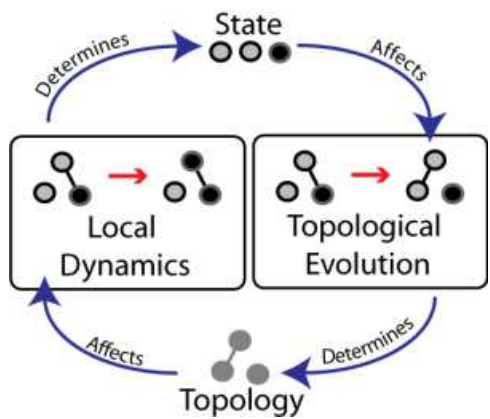}
\caption{\label{figIllustration} (Color Online) In an adaptive network the evolution of the topology depends on the dynamics of the nodes. Thus a feedback loop is created in which a dynamical exchange of information is possible}
\end{figure}

A network consists of a number of network \emph{nodes} connected by \emph{links} (see also Box 1).
The specific pattern of connections defines the network's \emph{topology}.
For many applications it is not necessary to capture the topology of a given real world network exactly in a model. Rather, the processes of interest depend in many cases only on certain \emph{topological properties} \cite{Costas:Measures}.
The majority of recent studies revolve around two key questions corresponding to two distinct lines of research: what are the values of important topological properties of a network that is evolving in time and, secondly, how does the functioning of the network depend on these properties?   

The first line of research is concerned with the \emph{dynamics of networks}. Here, the topology of 
the network itself is regarded as a dynamical system. It changes in time according to specific, often local, rules. 
Investigations in this area have revealed that certain evolution rules give rise to peculiar network topologies 
with special properties. Notable examples include the formation of small world \cite{Watts:SmallWorlds} 
and scale free networks \cite{Price:PowerLaw,Barabasi:Scaling}. 

The second major line of network research focuses on the \emph{dynamics on networks}.
Here, each node of the network represents a dynamical system. The individual systems are 
coupled according to the network topology. Thus, the topology of the network remains static while
the states of the nodes change dynamically. Important processes that are studied within this framework 
include synchronization of the individual dynamical systems \cite{Pecora:SmallWorldSync} and contact processes, 
such as opinion formation and epidemic spreading 
\cite{May:ScaleFree,PastorSantorras:ScaleFree,Kuperman:SmallWorldEpidemics,Newman:Assortative,PastorSantorras:NoThresholds}. 
These studies have made it clear that certain topological properties have a strong impact on 
the dynamics. For instance it was shown that vaccination of a fraction of 
the nodes cannot stop epidemics on a scale free network \cite{PastorSantorras:ScaleFree,May:ScaleFree}.

Until recently, the two lines of network research were pursued almost
independently in the physical literature. While there was certainly a strong interaction 
and cross-fertilization, a given model would either describe the dynamics \emph{of} a 
certain network or the dynamics \emph{on} a certain network. Nevertheless, it is clear that in most 
real world networks the evolution of the topology is invariably linked to the state of the 
network and vice versa. 
Consider for instance a road network. The topology of the network, that is the 
pattern of roads, influences the dynamic state, i.e. the flow and density of traffic. But, if traffic congestions 
are common on a given road, it is likely that new roads will be build in order to decrease 
the load on the congested one. In this way a feedback loop between the state and topology of the network is formed. 
This feedback loop can give rise to a complicated mutual interaction between a time varying network topology and
the nodes' dynamics. Networks which exhibit such a feedback loop are called 
\emph{coevolutionary} or \emph{adaptive} networks (s. also Fig.~\ref{figIllustration}).

Based on the successes of the two lines of research mentioned earlier, it is the next
logical step to bring these strands back together and to investigate the dynamics of
adaptive networks.
Indeed, a number of papers on the dynamics of adaptive networks have recently appeared. Since adaptive networks occur over a large variety of scientific disciplines they are currently investigated from many different directions. 
While the present 
studies can only be considered as first steps toward a theory of adaptive networks, they already 
crystallise certain general insights. Despite the thematic diversification, the reported results, 
considered together, show that certain dynamical phenomena repeatedly appear in adaptive networks:
the formation of complex topologies, robust dynamical self-organization, spontaneous emergence 
of different classes of nodes from an initially inhomogeneous population, and complex mutual dynamics in state 
and topology. In the following we argue that the mechanisms that give rise to these phenomena arise from 
the dynamical interplay between state and topology. They are therefore genuine adaptive network effects that
cannot be observed in non-adaptive networks. 

In this review it is our aim to point out that many recent findings reported mainly in the physical 
literature describe generic dynamical properties of adaptive networks. These findings are therefore of 
potential importance in many fields of research. In particular we aim to make recent insights accessible to researchers in the biological sciences, where adaptive networks frequently appear and have been studied implicitly for a long time.    

We start in Sec.~\ref{secApplied} by discussing several examples that illustrate the abundance of 
adaptive networks in the real world and in applied models. Thereafter we proceed 
to the core of the review. In Sec.~\ref{secBoolean} adaptive Boolean networks are studied
to explain how adaptive networks can self-organize towards dynamical criticality. 
Other, less obvious, but no less intriguing forms of the self-organization
are discussed in Sec.~\ref{secWeighted} while we review investigations of adaptive
coupled map lattices. In particular, it is shown that a spontaneous `division of labour'
can be observed in which the nodes differentiate into separate classes, which play
distinct functional roles in the network. Further examples of this functional
differentiation of nodes are discussed in Sec.~\ref{secGames}, which focuses on games 
on adaptive networks. Finally, in Sec.~\ref{secSpreading} we discuss  the dynamics of
the spreading of opinions and diseases on social networks, which shows that the adaptive
networks can exhibit complex dynamics and can give rise to new phase transitions. We
conclude this review in Sec.~\ref{secSummary} with a summary, synthesis and outlook. 


\section{Ubiquity of adaptive networks across disciplines\label{secApplied}}
Adaptive networks arise naturally in many different applications. Although studies that target the interplay between network state and topology have only recently begun to appear, models containing adaptive networks have a long tradition in several scientific disciplines. 
In the introduction we have already mentioned the example of a road network that can be considered as a prototypical adaptive network. Certainly, the same holds for many other technical distribution networks such as power grids \cite{Scire:PowerGrid}, the mail network, the internet or wireless communication networks \cite{Greiner:MultiHop,Krause:Wireless,BarYam:Wireless}. In all these systems 
a high load on a given component can cause component failures, e.g. traffic jams or electrical line failures, with the potential to cut links or remove nodes from the network. On a longer timescale, high load will be an incentive for the installation of additional connections to relieve this load. 
Further, games on adaptive network have recently become a hot topic in the engineering literature where they are called \emph{network creation games}. These are currently investigated in the context of evolutionary engineering\cite[and references therein]{Greiner:Engineering}.  

Essentially the same mechanisms are known to arise in natural and biological distribution networks. Consider for example, the vascular system. 
While the topology in the network of blood vessels directly controls the dynamics of blood flow, the blood flow also exhibits a dynamic feedback on the topology.
One such process is arteriogenesis, where new arteries are formed to prevent a dangerous restriction in blood supply (ischemia) in neighbouring tissues \cite{Schaper:Arteriogenesis}.

More examples of adaptive networks are found in information networks like neural or genetic 
networks. In the training of an artificial neuronal network, for example, it is obvious that 
the strength of connections and therefore the topology has to be modified depending on 
the state of the nodes. The changed topology then determines the dynamics of the state in the 
next trial. Also in biological neural and genetic networks some evidence suggests that the 
evolution of the topology depends on the dynamics of the nodes \cite{Hopfield:Unlearning}. 

In the social sciences networks of relationships between individuals or groups of
individuals have been studied for decades. On the one hand important processes like
the spreading of rumours, opinions and ideas take place on social networks -- and are influenced 
by the topological properties. On the other hand it is obvious that, say political opinions or religious beliefs, can in turn have an impact on the topology, when for instance conflicting views lead to the breakup of social contacts. 

In game theory there is a long tradition to study the evolution of cooperation in 
simple agent-based models. In recent years spatial games that are played 
on social networks have become very popular. While most studies 
in this area so far focus on static networks, one can easily imagine that the 
willingness of an agent to cooperate has an impact on his social contacts. 
To our knowledge the huge potential of games on adaptive networks was first pointed 
out by Skyrms and Pemantle\cite{Skyrms:Social}. 

Further examples of adaptive networks are found in chemistry and biology. 
A model of an adaptive chemical network originally proposed by Jain and Krishna is studied in 
\cite{Jain:Cooperation,Schweitzer:Aggregate}. In the model the nodes of the networks are chemical 
which interact through catalytic reactions. Once the population dynamics has reached an
attractor the species with the lowest concentration is replaced by a new species with
randomly generated interactions. Although the topology of the evolving network is not
studied in great detail, this paper shows that the appearance of a topological
feature--an autocatalytic loop--has a  strong impact on the dynamics of both state and
topology of the network. 

While Jain and Krishna focus on the evolution of chemical species, their work is clearly 
inspired by models of biological evolution. In ecological research models involving
adaptive networks have a long tradition. A prominent area in which adaptive networks
appear is food web evolution \cite{Drossel:webworld,Dieckmann:AdaptiveSpeciation,Dieckmann:Nature}. 
Food webs describe communities of different populations
that interact by predation. In almost all models the 
abundance of a species, i.e. the dynamic state,
depends on the available prey
as well as on the predation pressure, both of which depend in turn on topology of the
network. If the population size drops below a certain threshold the corresponding population
goes extinct and the node is removed from the network. Therefore the
dynamics of the topology depends on the state of the network.

The examples discussed above show that adaptive networks appear in a large variety of
different contexts. However, the nature and dynamics of the adaptive feedback as such has
to-date only been investigated in a relatively small number of studies. In the following sections
we focus on papers that specifically investigate the adaptive interplay of
state and topology and illustrate the implications this interplay can have. 

A prominent ancestor of research in adaptive networks is \cite{Christensen:AdaptiveBakSneppen}.
In this work Christensen et al.~discuss a variant of the famous Bak-Sneppen model of
macro-evolution \cite{Bak:SneppenModel}. The model describes the evolution of a number of populations, represented as nodes of a network in which the
(undirected) links correspond to abstract ecological interactions. The state of each node
is a scalar variable that denotes the population's evolutionary fitness. Initially this fitness is assigned randomly.  Thereafter, the model is updated
successively by replacing the population with the lowest fitness by a new species with
random fitness.  The replacement of a species is assumed to affect also the fitness of
the other populations it is interacting with. Therefore, the fitness of all neighbouring species
(that is, species with direct links to the replaced one) are also set to random values.
In the original model of Bak and Sneppen the underlying network is a one dimensional
chain with periodic boundary conditions, so that every node has exactly two
neighbouring nodes. In other words, the \emph{degree} of each node is two. 
It is well known that this model gives rise to avalanches of
species replacements which follow a scale free size distribution \cite{Bak:SneppenModel}.

\begin{figure}[htb]
\includegraphics[width=3in]{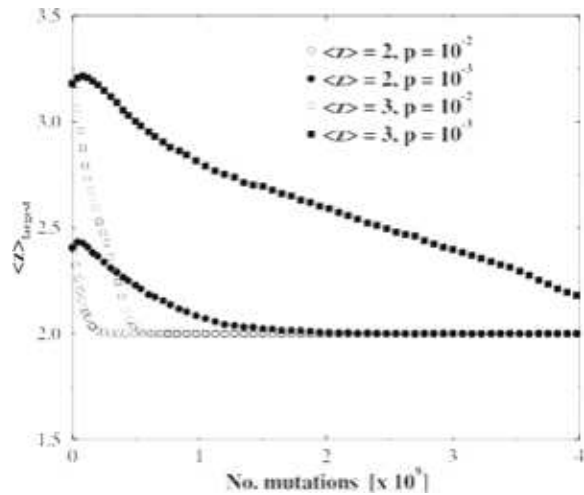}
\caption{For different initial conditions the connectivity $\langle z\rangle_{\rm largest}$
of the largest connected cluster, of the adaptive boolean network studied by Christensen et 
al. self organizes towards the critical value of $2$. Source: K.~Christensen et al.~
Phys.~Rev.~Lett.~{\textbf 81} (11), 1998 \cite{Christensen:AdaptiveBakSneppen} Fig.~3.)
\label{figChristensen}}
\end{figure}

In the paper by Christiansen et al.~the simple topology of the Bak-Sneppen model is replaced
by a random graph \cite{Christensen:AdaptiveBakSneppen}. The paper focuses mainly on
the evolutionary dynamics on networks with static topology. However, in the second to last paragraph
a model variant is studied in which the replacement of a population can affect the local
topology. If the replaced population had a lower degree than the species in the
neighbourhood, there is a small probability that a new link is added that connects to the
replaced species. But, if the replaced population had a higher degree than the species
in the neighbourhood than one link that connects to the replaced species is removed with
the same probability. This evolution rule effectively changes the \emph{mean degree}, that is the average number of links connecting to a node. By numerical simulation Christensen et 
al.~find that the mean degree in the largest cluster of nodes approaches two (see~Fig.~\ref{figChristensen}) -- exactly the same mean degree as the linear chain used in the original Bak-Sneppen model.
This finding is remarkable since it suggests that adaptive networks are capable of robust
self-organization of their topology based on local rules. This observation triggered a number of subsequent
studies which will be discussed in the next section. 

\section{Robust self-organization in boolean networks \label{secBoolean}}
In order to understand the functioning of adaptive networks it is reasonable to focus on
conceptually simple models. In Boolean networks the state of a given node is characterized 
by a single Boolean variable. Boolean networks with variable topology, offer a particularly simple and well-studied framework for the study of dynamical phenomena. Two prominent applications of Boolean networks are neural and gene regulatory nets, in which the state of a given node indicates whether a certain gene is being transcribed or whether a certain neuron is firing. 

It is known that Boolean networks are capable of different types of dynamical behaviour,
including chaotic and stationary (frozen) dynamics \cite{Kauffman:Networks}. At the boundary between stationarity and chaos, lies often a narrow transition region, where oscillatory dynamics can be observed and the density of frozen nodes exhibits power-law scaling. 
According to biological reasoning, neural as well as gene regulatory networks have to be close to
or on this `edge of chaos' to function properly (e.g., to code for different distinct
cell types or allow meaningful information processing). A central question is how the
networks manage to stay in this narrow parameter region while undergoing topological changes 
in the course of biological evolution and individual development. As we will show in the following 
it is likely that the adaptive nature of these networks plays a central role in the self-organization 
towards the critical oscillatory or quasi-periodic states.

Perhaps the simplest models for regulatory and neural nets are \emph{threshold networks}. 
In these networks the state of the Boolean variable indicates whether the corresponding node is 
\emph{active} or \emph{inactive}. Depending on the 
topology active nodes may exert a promoting or an inhibiting influence on their direct neighbours in the network. If the inputs received by a node exceed 
a certain threshold, say if a node receives more promoting than inhibiting signals via its links, 
the node becomes active; otherwise it is inactive.   

In order to study topological self-organization Bornholdt and Rohlf \cite{Bornholdt:AdaptiveBoolean} used 
a Boolean threshold network in conjunction with an update rule for the topology:
The time evolution of the system is simulated until a dynamical attractor, say a limit cycle, 
has been reached. Then a randomly chosen node is monitored for one period of the attractor or, 
in case of chaotic dynamics, for a long fixed time. If the state of the node changes at least once
during this time it loses a random link. However, if the state remains 
unchanged for the whole duration, a link from a randomly selected node is created. 
In short, `frozen' nodes grow links, while `dynamical' nodes loose links.

Note that adding links randomly can lead to the formation of, apparently non-local, long 
distance connections. However, since the targets of the links are randomly determined no distributed 
information is used. In this sense topological evolution rules that add or remove 
random links can be considered as local rules. 

By numerical simulation Bornholdt and Rohlf show that, independently of the initial state, 
a certain level of connectivity is approached. If the number of nodes $N$ is changed the emerging 
connectivity $K$ follows the power law $K=2+12.4 N^{-0.47}$. Therefore, in the case of large 
networks ($N\to\infty$) self-organization towards the critical connectivity $K_c=2$ can be observed. 
This is explained by further simulations which show that in large networks a topological phase 
transition takes place at $K=2$. In this transition the fraction of `frozen' nodes
drops from one to zero: Before the transition all nodes change their state in one period of the attractor, while above the transition almost no node changes its state at all. This means, in a 
large network, the proposed rewiring algorithm almost always adds links if $K<2$, but almost always 
removes links if $K>2$. In this way self-organization towards the dynamically critical state 
takes place. This form of self-organization is highly robust as it does not depend sensitively on the 
initial topology or the choice of parameters.

As is pointed out in \cite{Bornholdt:AdaptiveBoolean} and later in a different context in \cite{Bornholdt:Neural}
these results illustrate an important principle: Dynamics on a network can make information 
about global topological properties locally accessible. In an adaptive network this 
information can feed back into the local dynamics of the topology. Therefore, the adaptive interplay between 
the network state and topology can give rise to a highly robust global self-organization based 
on simple local rules. Note, that this genuine adaptive network effect can  be observed in 
networks where topological evolution and dynamics of the states take place on separate time scales, 
as the example presented by Bornholdt and Rohlf shows. These results inspired several subsequent 
investigations that extended the results
\cite{Bornholdt:RobustnessEvolution,Bornholdt:Genetic,Luque:Boolean,Bornholdt:SelfOrganizedMedia,Bornholdt:Neural,Liu:Boolean,Rohlf:Threshold}. 

A natural generalization of the system of Bornholdt and Rohlf is to replace 
the threshold function by more general Boolean functions.
In the Kauffman networks studied in \cite{Bornholdt:Genetic}, \cite{Liu:Boolean} and \cite{Luque:Boolean} random 
Boolean functions are used, which are represented by randomly created lookup tables. 
In \cite{Luque:Boolean} these lookup tables are created with a bias $p$ so that a random input leads to 
activation with probability $p$ and deactivation with probability $1-p$. In this way networks 
are created in which the critical connectivity can be tuned by changing $p$.
Although a different rewiring rule is used, only allowing for disconnection, 
self-organization of the system towards the critical state (from above) is still observed.

The work described above shows that already very simple adaptive networks can exhibit complex 
dynamics. In order to find further examples of sets of interesting rules an exhaustive search over a large class of adaptive network models is desirable. Indeed, first attempts in this direction 
for Boolean networks have been reported in \cite{Sayama:GNA}. In particular, a numbering scheme is proposed that allows to enumerate all adaptive networks in a given class. A similar formal, 
cellular-automaton-inspired approach is also used in \cite{Smith:FunctionalNetwork}.  

Finally let us remark, that beside the mechanism described by Bornholdt and others there exists an alternative mechanism for making information on the global state locally available, which again can be utilized to robustly self organize the system. 
This `dual' mechanism applies if the topology of the network changes much faster than the state. For illustration consider the following toy model: 
In a given network links are established randomly, but links between nodes of different states are instantaneously broken. These rules lead to a configuration in which every node of a given state is connected 
to all other nodes in the same state. This means that if a given node has, say, five links there are exactly five other nodes in the network that have the same state. Global information on the states has become locally available through the topology. This information can now feed back into the dynamics of the states on a slower timescale.


\section{Leadership in coupled oscillator networks\label{secWeighted}}
In the previous section we have discussed the adaptive interplay between state and topology 
as a dynamical feedback that can drive systems towards criticality.
A similar feedback loop can, in a slightly different setting, guide the 
self-organisation towards non-trivial topologies. One possible outcome 
is a spontaneous `division of labour': the emergence of distinct classes of nodes
from an initially homogeneous population. This phenomenon was first described by Ito and Kaneko 
\cite{Ito:Leaders, Ito:GCM} in an adaptive network of coupled oscillators. It is remarkable that 
these authors state with great clarity, that their work was motivated by the new dynamical 
phenomena that can be expected in adaptive networks. 

Ito and Kaneko study a directed network, in which each node represents a chaotic oscillator.
The state of an oscillator is characterized by a continuous variable. Furthermore, every 
link in the network has an associated continuous variable that describes its \emph{weight}, 
that is the strength of the connection. The states of the nodes and the weights of the links are 
updated in discrete time steps. In these updates the new state of a given node is determined 
according to a logistic map \cite{May:Logistic} which is coupled to the neighbouring oscillators via 
the network connections. The weights follow an update rule which increases the coupling between 
oscillators in similar states, while keeping the total weights of all inputs to each single oscillator 
constant.    
 
The use of weighted networks is a convenient choice for the analysis of structural changes in adaptive networks. For example, they can be initialized with uniform weights and states plus minor fluctuations. Effectively that means that all oscillators are initially in almost 
identical states and are connected to all other oscillators with equal strength. That is, initially the nodes form a homogeneous population. However, over the course of the simulation the weight 
of a large fraction of links approaches zero, so that a distinct network structure emerges. This structure 
can be visualized (and analyzed) by only considering links above a certain weight and neglecting 
all others. In the model of Ito and Kaneko this network does not approach a frozen configuration, but 
remains evolving as links keep appearing and disappearing by gaining or losing weight.  

Ito and Kaneko show that in a certain parameter region two distinct classes of nodes form that differ by 
their effective outgoing degree.  
Even a network in which some nodes are of high degree while other nodes are of low degree could still be considered to be homogeneous \emph{on average} if every node has a high degree at certain times and a low outgoing degree at others. However, in the model of Ito and Kaneko this is not the case: Despite the ongoing rewiring of individual links, a node that has a high/low outgoing degree at some point in time will generally have a high/low outgoing degree also later in time. Note that the outgoing degree indicates the impact that a given node has on the dynamics of others in the network. In this sense one could describe the findings of Ito and Kaneko as the emergence of a class of `leaders' and a class of `followers' - or, to use a more neutral metaphor, of a spontaneous `division of labour' in which the nodes differentiate to assume different functional roles.   

A similar division of labour was subsequently observed in a number of related systems
which can be interpreted as simple models of neural networks \cite{Leeuwen:SmallWorld,Leeuwen:FollowUp,Bornholdt:AdaptiveBoolean}.  
As a common theme, in all these models the topological change arises through a strengthening of connections between elements in a similar state -- a rule that is for neural networks well motivated by empirical results \cite{Paulsen:Plasticity}. 
Even in systems in which no distinct classes of nodes emerge the strengthening of connections 
between similar nodes often leads to strong heterogeneity in degree. A notable example is the 
formation of a scale free topology reported in \cite{Fan:Growth} and \cite{Fronczak:Sandpile}.  

From a technical point of view the emergence of strong heterogeneity in degree is not always 
desirable. For instance it is known that homogeneous networks, consisting of nodes with a similar degree, are more easy to synchronize \cite{Donetti:EntangledWebs}. In an interesting 
paper Zhou and Kurths \cite{Zhou:AdaptiveNets} study an adaptive network of coupled 
chaotic oscillators in which the connections between different nodes are strengthened.
Note that this is exactly the opposite of the adaptation rule proposed by Ito and Kaneko. Consequently, the adaptive self-organization drives the network into the direction of a more homogeneous topology, ongoing with an enhanced ability for synchronization. Thereby it is possible to synchronize networks that exceed by several orders of magnitude the size of the largest comparable random graph that is still synchronisable.

\begin{figure}
\includegraphics[width=3in]{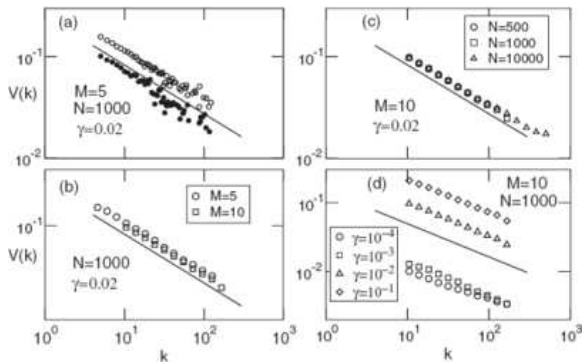}
\caption{The adaptive network of coupled oscillators studied by Zhou and Kurths 
organises towards a topology in which the incoming weight 
$V_i$ is a power law of the nodes degree $k$. The exponent $\theta=-0.48$ is 
independent of (a) the specific type of oscillator under consideration,
(b) the mean degree $M$, (c) the size of the network and (d) the adaptation parameter $\gamma$. 
Source: Zhou and Kurths, Phys.~Rev.~Lett.~{\textbf 96}, 164102, 2006 \cite{Zhou:AdaptiveNets} Fig.~2.  
\label{figZhou}}
\end{figure}

Another hallmark of adaptive networks that reappears in the work of Zhou and Kurths is 
the emergence of power laws. They show that in the synchronized state the incoming 
connection weights $V_i$ scale with the degree $k$ of the corresponding node $x_i$ as
$V(k)\sim k^{-\theta}$, where $\theta=-0.48$ is independent of the parameters in the model (see Fig.~\ref{figZhou}). 
The authors point out that this universal behavior arises because of a hierarchical transition to synchronization.
In this transition the nodes of the highest degree are synchronized first. Nodes of lower degree 
are synchronized later and therefore experience the increase in coupling strength for a longer time.


Let us remark that the results reported in this section indicate that there could be a subtle 
connection to the mechanism described by Bornholdt and Rohlf \cite{Bornholdt:AdaptiveBoolean}: The results of 
Ito and Kaneko show that there is a scale separation between the dynamics of the network 
(involving states and topology) and the timescale on which the \emph{emergent} properties 
of the nodes change. In other words the turnover time for a node of high degree 
to become a node of low degree is many orders of magnitude larger than the time required for 
the rewiring of individual links. In contrast to other models this time scale separation 
is not evident in the rules of the system but emerges from the dynamics. One can suspect that this time 
scale separation could arise because of the presence of a phase transition at which the turnover 
time diverges. In the light of the findings described in the previous section it is conceivable 
that an adaptive network could self-organize towards such a phase transition. However, more 
investigation in this direction are certainly necessary to verify that this is indeed the case. 


\section{Cooperation in games on adaptive networks \label{secGames}}
The term `division of labour' used in the previous section already suggests a socio-economic reading. 
Indeed, socio-economic models are perhaps the most fascinating application of adaptive networks so far. 
In this context the nodes represent agents (individuals, companies, nations, \ldots{}) while the links 
represent social contacts or, say, business relations. In contrast to other systems considered so 
far agents are in general capable of introspection and planning. For this reason the exploration 
of socio-economic systems is invariably linked to game theory. 

One of the central questions in game theory is how cooperation arises in populations
despite the fact that cooperative behaviour is often costly to the individual. A paradigmatic game which 
describes advantageous but costly cooperation is the \emph{prisoner's dilemma}. In this game
two players simultaneous chose between cooperation and defection. From the perspective of a 
single player choosing to defect always yields a higher payoff regardless of the action 
of the opponent. However, the collective payoff received by both players is the lowest if both players defect and the highest if both cooperate.

In models, the action a player takes is determined by its strategy, which comprises of
a lookup table, that maps the information 
from a given number of previous steps to an action,
as well as rules for the initial rounds where no such information is available. By updating 
the strategies of players according to a set of evolutionary rules, the evolution of cooperation  can be studied. 

Spatial games in which the players are arranged on a static network with links that 
represent possible games have been studied for some time (e.g. \cite{Novak:SpatialGames}).
More recently games on adaptive networks have come into focus. In these games the 
players can improve their topological position, for example by cutting links 
to defectors. 
The prisoner dilemma game on adaptive networks has been studied 
in \cite{Bornholdt:AdaptiveGames2, Zimmermann:Prisoner, Zimmermann:Differentiation, Zimmermann:Leaders, Traulsen:Mapping}. 
An adaptive version of the closely related Snowdrift game was investigated in \cite{Ren:Snowdrift} and a more realistic socio-economic model involving taxes and subsidies was discussed in \cite{Lugo:Taxes}.
In the results presented in these papers the two common themes discussed above,
namely the robust topological self-organization and the associated appearance of power laws,  reappear and have been noted by many authors. For instance the formation of scale-free topologies, which exhibit a power law degree distribution, is discussed in detail by Ren et al. \cite{Ren:Snowdrift}
and Egu\'{\i}luz et al. \cite{Zimmermann:Differentiation}. 

An effect that is reminiscent of the spontaneous division of labour and the emergence of social hierarchies 
was observed by Zimmermann et al.~\cite{Zimmermann:Prisoner}, Egu\'{\i}luz et al.~\cite{Zimmermann:Differentiation} and Zimmermann and Egu\'{\i}luz\cite{Zimmermann:Leaders}. However, in these paper the adaptive interplay between the 
network state and topology stops at some point as the network freezes in a final configuration, a so-called \emph{network Nash equilibrium}.  It is therefore not clear whether the different 
social classes observed in the simulations arise because of the same mechanism as in the model of Ito and Kaneko. As another possible explanation the network could have reached an absorbing state, freezing the network and thus fixing local topological heterogeneities in some otherwise transient state.

An observation reported by Ebel and Bornholdt \cite{Bornholdt:AdaptiveGames2} as well as 
Egu\'{\i}luz et al.~\cite{Zimmermann:Differentiation} and Zimmermann and Egu\'{\i}luz~\cite{Zimmermann:Leaders}
is that the approach to the final state is marked by large avalanches of strategy changes which 
exhibit power-law scaling. Such scaling behaviour is another indicator of self-organized critical 
behaviour.

From an applied perspective it is interesting that elevated levels of cooperation 
are reported in all papers cited above. The mechanism that promotes cooperation in adaptive networks 
becomes apparent when one considers the interaction between the players and their neighbourhood. 
In all games on networks the local neighbourhood acts as an infrastructure or substrate from 
which payoffs are extracted. The quality of this infrastructure depends on topological properties 
such as the degree or the number of cooperators in the neighbourhood. In an adaptive network 
a player can shape this neighbourhood by its own actions. Thereby the neighbourhood becomes 
an important resource. The rules of the games are generally such that selfish behaviour 
degrades the quality of this resource as neighbours cut or rewire their links. This feedback 
may be regarded as a `topological punishment' of the defecting player. 

A rigorous investigation in the mechanism that promotes cooperation on adaptive networks 
is presented by Pacheco et al.~\cite{Traulsen:Mapping}. In the limit in which topological 
dynamics is much faster than the evolution of strategies the authors show that the prisoner 
dilemma on an adaptive network can be mapped to a game in a well mixed population. However, 
this `renormalized' game is not a prisoner dilemma; the mapping effectively changes the rules 
of the game so that the prisoner dilemma is transformed into 
a coordination game. This explains the elevated levels of cooperation since the cooperative 
behaviour is naturally favoured in the coordination game.     

It is interesting to note that the adaptive nature of a network is not always 
apparent on the first glance. For instance Paczuski et al.~\cite{Bassler:Boolean} study the minority 
game on a fixed network. In this non-cooperative game each agent makes a decision 
between two alternatives. The agents who decide for the alternative chosen by the 
minority of agents are rewarded. The decision of the agent depends on its 
own decision in the previous round as well as on the decision of its immediate neighbours 
in the network during that round. As in the prisoner dilemma the strategy
of an individual agent can be described by a lookup table that is allowed to evolve in time to maximize success. Despite the fact that the game is seemingly played on a static network 
Paczuski et al.~observe all the hallmarks of adaptive networks described above, including the emergence of two distinct groups which differ in their success in the game. This enigma is resolved by noting that the evolution of the strategies in the lookup tables can effectively change the nature of the links in the network. In particular the lookup tables can evolve 
to such a state that the decision of certain neighbours in the network is ignored 
entirely \cite{Maya}. This means that even though the network itself is static the effective degree, which is experienced by the nodes, can change over time. Therefore the network is after all adaptive.     

While adaptive networks can add realism to previously studied games like the prisoner's 
dilemma, they also give rise to a new class of games. In these games the players 
do not try to maximize an abstract payoff, but struggle to achieve an advantageous 
position on the network. For example, in a social network a position of high 
centrality is certainly desirable. The struggle for such a position is studied in models 
by Rosvall and Sneppen \cite{Sneppen:SelfAssembly,Sneppen:AdaptiveNetworks,Sneppen:Structures}. The model 
describes the formation of a communication network between social agents. As an interesting feature of this model the communication provides the agents with metainformation about the 
network structure. 

\begin{figure}[tb]
\includegraphics[width=3in]{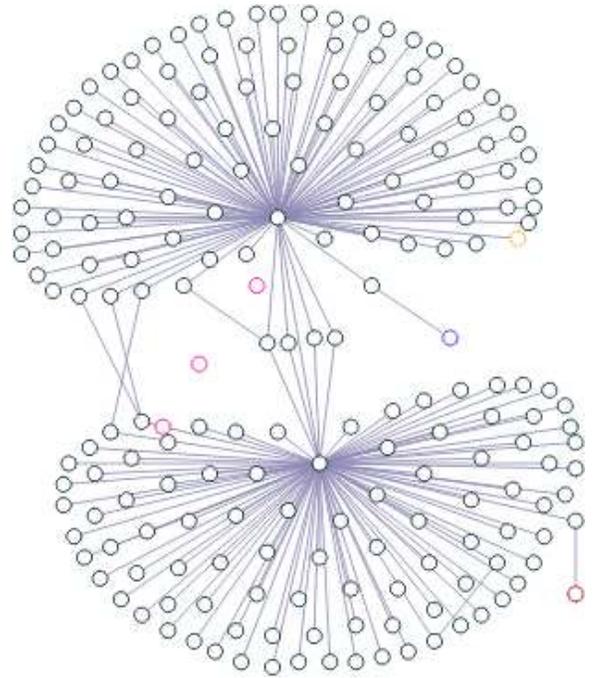}
\caption{(Color Online) In the paper of Holme and Ghoshal, agents compete for a position of high centrality 
and low degree. This figure shows that complex global topologies are formed. In the figure 
three classes of nodes can be identified. Most nodes suffer from a low centrality, while 
others gain high centrality at the cost of having to maintain a large number of links. 
Only a small class of `VIP' nodes manage to achieve both high centrality and low degree. 
Source: Holme and Ghoshal, Phys.~Rev.~Lett.~{\bf 96}, 098701, 2006 \cite{Holme:NetworkAgents} Fig.~2b.  
\label{figHolme}}
\end{figure}

In a related model by Holme and Ghoshal \cite{Holme:NetworkAgents}
the agents attempt to achieve a position of high centrality while minimizing the 
number of contacts they have to maintain. 

Holme and Ghoshal show in simulations that the system exhibits long periods of stability 
where one strategy is dominant. These are interrupted by sudden invasions of a different strategy. 
Apparently, no steady state is approached so that the successional replacement of the 
dominant strategy continues in the long term behavior. An interesting feature of the model is that it transiently gives rise to highly nontrivial topologies. Figure \ref{figHolme} shows an example 
of such a topology. The shown topology is complex in the sense that it is immediately evident 
that it is not random or regular, but possesses a distinct structure. Note, that three distinct classes of nodes can be recognized in the figure. In particular there is a class of agents who achieve the goal of being 
in a position of high centrality and low degree. However, while a spontaneous division of labour
is evident, there is no de-mixing of classes: A node holding a position of low degree and high centrality at a certain time does not have an increased probability of holding such a position at a later time. Note also that the node's centrality that enters into the model is a global property. Therefore the emerging topologies are not organized based on local information alone.    


\section{Dynamics and phase transitions in opinion formation and epidemics\label{secSpreading}} 
Above we have mainly been concerned with systems in which the state of the network 
changes much faster or much slower than the evolution of the topology. In systems 
that exhibit such a time scale separation only the averaged state of the fast variables can affect the dynamics of the slow variables and therefore, the dynamical interplay between the time scales will, in general, be relatively weak. 
In contrast, new possibilities open up in systems in which the evolution of the topology takes place on the same timescale as the 
dynamics on the network. As dynamical variables and topological 
degrees of freedom are directly interacting a strong dynamical interplay between the 
state and topology becomes possible. One might say that information on the dynamics of the state can be stored in and read from the topology and vice versa. In the study of this interplay 
we can no longer make use of the time scale separation. Nevertheless it is still possible to analyze and understand the dynamics on the network by using the tools of nonlinear dynamics and statistical physics. Depending on the language of description the qualitative transitions in the dynamics and topology then become apparent in the form of either \emph{bifurcations} or \emph{phase transitions}.
   
A simple framework in which the dynamical interplay can be studied is offered by \emph{contact 
processes}, which describe the transmission of some property, e.g., information, political opinion, religious belief or epidemic infection along the network connections. One of the most simple models in this class is the epidemiological SIS model. This model describes a population of individuals forming a social network. Each individual is either susceptible (S) to the disease under consideration or infected (I). A susceptible individual in contact with an infected individual becomes infected with a fixed probability $p$ per unit time. Infected individuals recover at a rate $r$ immediately becoming susceptible again. If considered on a static network the SIS model has at most one dynamical transition. Below the transition only the disease-free state is stable, while above the transition the disease can invade the network and approaches an endemic state.  

\begin{figure}
\includegraphics[width=3in]{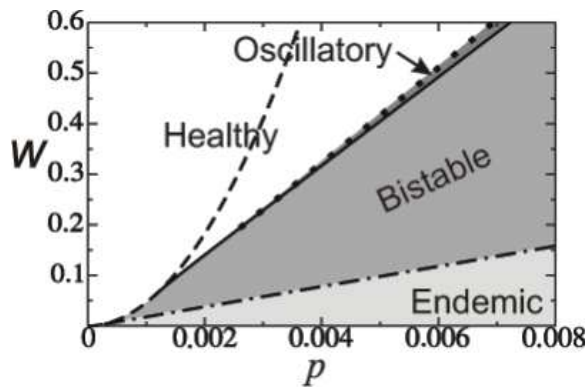}
\caption{Two parameter bifurcation diagram of the adaptive epidemiological network studied by Gross et al.
Bifurcations divide the parameter space in regions of qualitatively different dynamics. The 
dash-dotted line marks a transcritical bifurcation that corresponds to the threshold at which 
the epidemic can invade the disease free system. The region in which an established epidemic can remain 
in the system is bounded by a saddle-node bifurcation (dashed), a Hopf bifurcation (straight)
and a fold bifurcation of cycles (dotted).
Source: Gross et al., Phys.~Rev.~Lett.~{\textbf 96}, 208701, 2006 \cite{epidemicspaper} Fig.~4.  
\label{figGross}}
\end{figure}

The spatial SIS model can be turned into an adaptive network if an additional process is taken into account: Susceptible individuals can try to avoid contact with infected. Such a scenario was studied by Gross et al.~\cite{epidemicspaper}.
In their model with probability $w$ a given susceptible individual breaks the link to an infected neighbor and forms a new link to another susceptible. 
This additional rule turns the SIS model into an adaptive network. 
As was shown in Gross et al.~\cite{epidemicspaper}
even moderate rewiring probabilities change the dynamics of the system qualitatively. Sudden discontinuous transitions appear and a region of bistability emerges in which both the disease-free state and the epidemic state are stable (see~Fig.~\ref{figGross}). Similar findings are also reported by Ehrhardt et al.~\cite{Ehrhardt:AdaptiveSocial}
who investigate the spreading of innovation and similar phenomena on an adaptive network. 

At high rewiring rates the adaptive SIS model in \cite{epidemicspaper} can approach an oscillatory state in which the prevalence of the epidemic changes periodically. The oscillations are driven by two antagonistic effects of rewiring. On the one 
hand rewiring isolates the infected and thereby reduces the prevalence of the disease. On the other hand 
the rewiring leads to an accumulation of links between susceptibles and thereby forms a tightly connected 
cluster. At first the isolating effect dominates and the density of infected decreases. However, as the 
cluster of suscebtibles becomes larger and stronger connected a threshold is crossed at which the epidemic 
can spread through the cluster. This leads to a collapse of the susceptible cluster and an increased prevalence which 
completes the cycle. While this cycle exists only in a narrow region (Fig.~\ref{figGross}) in the model 
described above, the parameter region in which the oscillations occur and the amplitude of the oscillations 
are enlarged if one takes into account that the rewiring rate can depend on the awareness of the population 
and therefore on the prevalence of the epidemic \cite{epicoarse}.   

In the adaptive SIS model the hallmarks of adaptive networks discussed above reappear: The isolation of 
infected and the emergence of a single tightly connected cluster of susceptibles is an example of the 
appearence of global structure from local rules. Moreover, the mechanism that drives the oscillations 
is reminiscent of the self-organization to criticality discussed in Sec.~\ref{secBoolean}. 

\begin{figure}
\includegraphics[width=1.5in]{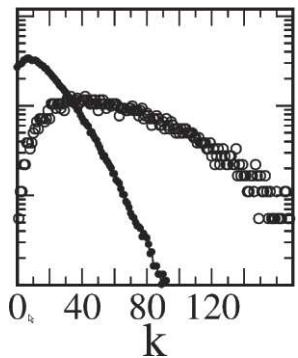}
\caption{\label{figGross2}In the model of Gross et al.~\cite{epidemicspaper} two topologically distinct populations of nodes emerge. These correspond to susceptibles (circles) and infected (dots) and have low and high degree $k$ respectively. }  
\end{figure}

The rewiring rule that is used in the adaptive SIS model establishes connections between nodes in identical 
states and severs connections between different states. Stated in this way the rewiring rule reminds of the 
model of Ito and Kaneko (see Sec.~\ref{secWeighted}) in which connections between similar nodes are strengthened and others weakened. This analogy suggests that topologically different classes of nodes 
could emerge from the dynamics of the network. Indeed, Fig.~\ref{figGross2} shows that two classes of 
nodes appear, which are characterized by different degree distributions. In this case we can identify the classes to consist of infected and of susceptible nodes, respectively.      

In order to study the dynamics of the adaptive SIS model Gross et al.~\cite{epidemicspaper}
and subsequently also Zanette \cite{Zanette:Closure} apply a moment closure approximation. 
By means of this approximation a low dimensional system of ordinary differential equations can be derived that captures the dynamics of the network. The system of equations can then be studied with the tools of bifurcation theory \cite{Guckenheimer:Dynamics,Kuznetsov:Elements} which reveal the critical points in parameter space where qualitative transitions in the dynamics occur. In order to capture the dynamics of the adaptive SIS model 
three dynamical variables are necessary, while the system-level dynamics of the standard (non-adaptive)
SIS model can be captured by only one variable. This shows that in the adaptive model two topological 
degrees of freedom communicate with the dynamics of the nodes.       

Another approach to the dynamics of adaptive networks is offered by the tools of statistical 
physics, which can reveal critical points in the form of phase transitions.
One example of such a phase transition is presented in a paper by Holme and Newman \cite{Newman:Social}, 
which focuses on opinion formation in populations. Specifically the paper considers the case 
of opinions, such as religious belief, for which the number of possible choices is only 
limited by the size of the population. Disagreeing neighbours manage to convince each other with 
probability $\phi$ or rewire their connections with probability $1-\phi$. This ultimately leads to a consensus 
state in which the network is decomposed into disconnected components, each of which consists
of individuals who hold a uniform opinion. For $\phi=0$ opinions never change, so that the final distribution 
of opinions matches the initial distribution. For $\phi=1$ no connections are rewired, so that 
the number of opinions in the consensus state can not exceed the number of disconnected components 
that already existed in the initial network. Applying a finite-size scaling analysis Newman and Holme are able to show that between these extremes a critical parameter value $\phi_c$ is located, at which 
a continuous phase transition takes place. At this transition a critical slowing down is observed, 
so that the network needs a particularly long time to reach the consensus state. In the consensus 
state the distribution of followers among the different beliefs approaches a power-law. 

The phase transition identified by Holme and Newman probably holds the 
key to the findings reported in \cite{Gil:AdaptiveOpinions,Zanette:FollowUp}. In this paper Gil and Zanette 
investigate a closely related model for the competition between two conflicting 
opinions. In this case conflicts are settled by 
convincing neighbours or cutting links. It is shown that a critical point exists 
at which only very few links survive in the consensus state. Based on the previous results 
it can be suspected that this is a direct consequence of the 
critical slowing down close to the phase transition. 
In this region the long time that is needed to settle to the consensus state might result in a very small number of surviving links.

\section{Summary, Synthesis and Outlook\label{secSummary}}

In this paper we have reviewed a selection of recently proposed models for adaptive networks. These examples illustrate that adaptive networks arise in a large number of different areas including ecological and epidemiological systems; genetic, neuronal, immune networks; distribution and communication nets and social models. The functioning of adaptive networks is currently studied from very different perspectives including nonlinear dynamics, statistical physics, game theory and computer science.

Despite the diverse range of applications from which adaptive networks emerge, we have shown that there are a number of hallmarks of adaptive behaviour that recurrently appear:
\begin{itemize}
\item{Self-organisation towards critical behaviour.}
Adaptive networks are capable of self organising towards a dynamically critical states, such as phase transitions. This frequently goes together with the appearance of power-law distributions. 
Unlike other forms of self-organized criticality this mechanism is highly robust (see Sec. 3). 
\item{Spontaneous `division of labour'.} 
In adaptive networks classes of topologically and functionally distinct nodes can arise 
from an initially homogeneous population. In certain models a `de-mixing' of these classes 
is observed, so that nodes that are in a given class generally remain in the class (see Sec. 4). 
\item{Formation of complex topologies. }
Even very basic models of adaptive networks that are based on very simple local rules 
can give rise to complex global topologies (see Sec. 5). 
\item{Complex system-level dynamics. }
Since information can be stored and read from the topology, the dynamics of adaptive 
networks involves local as well as topological degrees of freedom. Therefore the dynamics 
of adaptive networks can be more complex than that of similar non-adaptive models (see Sec. 6). 
\end{itemize} 

In the context of biological applications, the hallmarks described above can be used 
as a working guideline: If one of these phenomena is observed in nature one should 
consider the possibility that it is caused by an (possibly so-far unobserved or not recognized) adaptive network. 
As was demonstrated in the example of Paczuski et al.~\cite{Bassler:Boolean} the adaptive nature of a network may not always be obvious, but it can be revealed by a direct search. 
The reverse approach can also be rewarding: In systems which are known to contain an adaptive network it is promising to search for the hallmarks described above.  

Given the evidence that is summarized in this review, we believe that adaptive 
networks could hold the key for addressing several current questions in many areas of research, but in particular 
in biology. Adaptive self-organization could explain how neural and genetic networks 
manage to exhibit dynamics that in many models only appears in critical states at the `edge 
of chaos'. Spontaneous division of labor could be 
important for many social phenomena, such as leadership in simple societies, but also 
for developmental problems such as cell differentiation in multicellular organisms. 
The capability of adaptive networks to form complex topologies 
has not been studied in much detail, but it seems to offer a highly elegant way to 
build up large-scale structures from simple building blocks. A biological example 
where this certainly plays the role is for instance the growth of vascular networks. 

Many important processes have so far mainly been studied only on static networks. However, by doing so important aspects of such systems may be overseen or neglected. Take for example the spread of infectious diseases. Currently huge efforts are made to determine the structure of real world social networks. These are then used as input into complicated prediction models, which help to forecast the spread and dynamics of future epidemics (e.g. influenza). However, the most involved model or the best survey of the actual social network is in vain if it is not considered that people may radically change their behaviour and social contacts during a major epidemic.

We want to stress that answers to the questions outlined above would not only enhance our understanding of real world systems comprising of adaptive networks, but could also be exploited in bio-inspired technical applications that self-assemble or self-organize many subunits towards desired configurations \cite[see for instance]{Kawamura:SelfHealing}. Such strategies are much sought for because many of these artificial systems will soon be too complicated to be easily designed by hand.
Thus adaptive network structures may hold the key to provide novel, much-needed design principles and could well radically change the way in which future electrical circuits, production systems or interacting swarms of robots are operating.

From an applied point of view it is desirable to compose an inventory of the types of microscopic 
dynamics that have been investigated in adaptive networks and their impact on system-level properties. 
Such an inventory could give researchers a guideline as to what kind of phenomena can be expected in natural systems where similar processes are at work. For instance we have seen that `like-and-like'
processes which strengthen connections between similar nodes quite universally seem to give rise 
to heterogeneous topologies and global structures. A rough sketch of such an inventory based on the 
papers reviewed here is shown in Box 2. In certain places the observations can be supplemented 
by mathematical insights. For instance in every scale separated system there has to be a discontinuous 
transition in the fast dynamics in order to maintain an adaptive interplay in the long term 
evolution of the system. Otherwise the fast dynamics is simply slaved to the slow dynamics.  
Nevertheless much more information on the dynamics of adaptive networks is necessary to fill the inventory. 
This information will most likely come from automated numerical studies of large classes of adaptive networks.

We note that the analysis of an adaptive network is not necessarily more involved than that of its static counterpart. While the nodes in static networks generally have different topological neighbourhoods, 
by contrast, the neighbourhood of nodes in adaptive networks changes over time. Because of this mixing of local topologies the network as such becomes more 
amendable to mean field descriptions. However caution is in order, because naive mean field approximations can fail if a spontaneous division of labour occurs in the system and is not taken into account.     

Apart from the investigation of more examples of adaptive networks more fundamental work is certainly necessary. 
The works reviewed in this paper can only be considered as a first step 
towards a theory of adaptive networks. However, some important principles 
are already beginning to crystallize. The mechanism that drives the robust self-organization 
towards criticality is quite well understood: The dynamics on the network makes 
topological degrees of freedom accessible in every node. It thus spreads information on topological properties 
across the network.
The local topological evolution can then react on this information and thus drive the topology to a topological phase transition 
at which the dynamics on the network is critical. Above we have conjectured that the 
observed `division of labour' could be driven by a similar mechanism, characterized 
by self-organization towards a phase transition at which the critical slowing down of the turnover times between emergent properties of nodes occurs. Moreover, the appearance of topologically distinct classes of nodes is certainly an important factor for the formation of complex topologies. Another factor is probably the dual mechanism described at the 
end of Sec.~\ref{secBoolean} by which global organization of the topology is possible. Finally, the investigations reported in Sec.~\ref{secSpreading} illustrate how topological degrees of freedom, acting as dynamical variables, can give rise to complex system-level dynamics. 
Thus, the four hallmarks described above seem after all to be connected. It is therefore not unlikely that all of these 
peculiar properties of adaptive networks can be explained by a single theory describing the transfer of information between state and topology of the network and the subtle interplay between different timescales.  

Since adaptive networks appear in many different fields and are already implicitly contained in many models a theory of adaptive networks can be expected to have a significant impact on several areas of active research. Future fundamental research in adaptive networks should focus on supplying and eventually assembling the building blocks for such a theory. While it has been shown that dynamics on the network can make global order parameters locally accessible, this mechanism has only been demonstrated for a few types of local dynamics. Except for these examples it is not clear which set of local rules reveals what kind of global information. 
Another open question is how exactly the observed `division of labour' arises and how exactly nontrivial global topologies emerge from the local interactions. Finally, it is an interesting question which topological 
properties are affected by a given set of evolution rules, so that they can act about topological degrees of freedom.  

While the study of adaptive networks is presently only a minor offshoot, the results summarized above lead us to believe that it has the potential to become a major new direction in network research. In particular the prospect of a unifying theory and the widespread applications highlight adaptive network as promising area for future investigations.

\acknowledgements{The authors would like to thank Luciano Costa, Maya Paczuski, Ira Schwartz and Arne Traulsen for valuable comments on an earlier draft of the manuscript.}

\bibliography{onnetworks,ofnetworks,networkreview,adaptivenetworks,mypapers}

\pagebreak

\begin{minipage}{3in}

BOX 1 -- A brief network glossary.
{
\footnotesize
\vspace{5mm}

{\bf Degree.} The degree of a node is the number of nearest neighbours to which a node is connected. The mean degree of the network is the mean of the individual degrees of all nodes in the network. \vspace{1mm}

{\bf Dynamics.} Depending on the context the term \emph{dynamics} is used in the literature to refer to a temporal change of either the state or the topology of a network. In this paper we use dynamics exclusively to describe a change in the state, while the term \emph{evolution} is used to describe a change in the topology. \vspace{1mm}

{\bf Evolution.} Depending on the context the term \emph{evolution} is used  in the literature to refer to a temporal change of either the state or the topology of a network. In this paper we use evolution exclusively to describe a change in the topology, while the term \emph{dynamics} is used to describe a change in the state. \vspace{1mm}

{\bf Frozen nodes.} A node is said to be frozen if their state does not change over in the long term behaviour of the network. In certain systems discussed here the state of frozen nodes can change nevertheless on an even longer (topological) time scale. \vspace{1mm}

{\bf Link.} A link is a connection between two nodes in the networks. Links are also sometimes called \emph{edges} or simply \emph{network connections}. \vspace{1mm}

{\bf Neighbours.} Two nodes are said to be neighbours if they are connected by a link. \vspace{1mm}

{\bf Node.} The node is the principal unit of the network. A network consists of a number of nodes connected by links. Notes are sometimes also called \emph{vertices}. \vspace{1mm}

{\bf State of the network.} Depending on the context the state of a network is used either to describe the state of the network nodes or the state of the whole network including the nodes and the topology. In this review we use the term \emph{state} to refer exclusively to the collective state of the nodes. Thus, the state is a-priori independent of the network topology. \vspace{1mm}

{\bf Topology of the network.} The topology of a network defines a specific pattern of connections between the network nodes. \vspace{1mm}
}
\end{minipage}

\begin{minipage}{3in}
BOX 2 -- A first rough attempt at an inventory of dynamics of adaptive networks
{
\footnotesize

{\bf Activity disconnects.}
Rule: Frozen nodes gain links, active nodes loose links. Outcome: Self organization 
towards percolation transition, active nodes scale as a power law. Examples: \cite{Bornholdt:AdaptiveBoolean,Rohlf:Threshold}
\vspace{1mm}

{\bf Like-and-like.} Rule: Connections between nodes in similar states are strengthened. 
Outcome: Heterogeneous topologies, possibly scale free networks, emergence of topologically 
distinct classes of nodes. Examples: \cite{Bornholdt:Neural,Ito:Leaders}  
\vspace{1mm}

{\bf Differences attract.}
Rule: Connections between nodes in different states are strengthened. Outcome: Homogeneous topologies, power-law distributed link weights. Example: \cite{Zhou:AdaptiveNets}
\vspace{1mm}
}
\end{minipage}

\pagebreak

\end{document}